\documentclass[twocolumn,showpacs,preprintnumbers,amsmath,amssymb]{revtex4}
\usepackage{amssymb}
\usepackage{graphicx}
\usepackage{dcolumn}
\usepackage{bm}
\usepackage{textcomp}

\newcommand\oprod[2]{\ensuremath{|#1\rangle\langle#2|}}

\begin{document}
\title{Decoy state method for measurement device independent quantum key distribution with different intensities in only one basis}
\author{Zong-Wen Yu$ ^{1,2}$, Yi-Heng Zhou$ ^{1}$,
and Xiang-Bin Wang$ ^{1,3\footnote{Email
Address: xbwang@mail.tsinghua.edu.cn}}$}

\affiliation{ \centerline{$^{1}$State Key Laboratory of Low
Dimensional Quantum Physics, Tsinghua University, Beijing 100084,
People¡¯s Republic of China}\centerline{$^{2}$Data Communication Science and Technology Research Institute, Beijing 100191, China}\centerline{$^{3}$ Shandong
Academy of Information and Communication Technology, Jinan 250101,
People¡¯s Republic of China}}

\begin{abstract}
We show that the three-intensity protocol for measurement device independent quantum key distribution (MDI QKD) can be done with different light intensities in {\em only one} basis. Given the fact that the exact values yields of single-photon pairs in the $X$ and $Z$ bases must be the same, if we have lower bound of the value in one basis, we can also use this as the lower bound in another basis. Since in the existing set-up for MDI-QKD, the yield of sources in different bases are normally different, therefore our method can improve the key rate drastically if we choose to only use the lower bound of yield of single-photon pairs in the advantageous basis. Moreover, since our proposal here uses fewer intensities of light, the probability of intensity mismatch will be smaller than the existing protocols do. This will further improve the advantage of our method. The advantage of using Z basis or X basis of our method is studied and significant improvement of key rates are numerically demonstrated.
\end{abstract}


\maketitle


 Device imperfection can cause serious problem in security of quantum key distribution (QKD)~\cite{BB84,GRTZ02}.  The major imperfection include the multi-photon events  of source and the limited signal detection rate of. The decoy state method~\cite{ILM,H03,wang05,LMC05,AYKI,haya,peng,wangyang,rep,njp} can help to make a set-up with an imperfect single photon source be as secure as that with a perfect single photon
source~\cite{PNS1,PNS}.

Besides the source imperfection, the limited detection rate is another big threaten to the security~\cite{lyderson}.
Theories of the device independent security proof~\cite{ind1} have been proposed
to overcome the problem. However, these theories are technical demanding.

Recently,  measurement device independent QKD (MDI-QKD) was proposed based on the idea of
entanglement swapping~\cite{ind3,ind2}.
If we want to obtain a higher key rate, we can choose to directly use an imperfect single-photon source such as the coherent state~\cite{ind2} with decoy-state method for this, say the MDI decoy-state method.  Calculation formulas for the practical decoy-state implementation with only a few different states has been studied in, e.g., Refs.~\cite{wangPRA2013,qing3}.

In the initial proposal\cite{ind2}, it is assumed to use different intensities calculate the lower bound values ($\underline{s_{11}^Z}$, $\underline{s_{11}^X}$) for yields of single-photon pairs in $Z$ basis and $X$ basis {\em separately}. Here in this work, based on the fact that the yields $s_{11}^{X}$ and $s_{11}^{Z}$ of single-photon pairs in the $X$ and $Z$ bases are equal, we propose to use differen intensities in one basis only. Since $s_{11}^Z=s_{11}^X$, the lower bound of one of them must be also the lower bound of another one. Given the experimental data in the normal case (linear loss for channel),
the calculated lower bound values for $s_{11}^X$ and $s_{11}^Z$ can be significantly {\em different}, but we can simply choose to use the bigger one. Therefore, we {\em only} need to use different intensities in
the basis that produce data for a larger lower bound value of yield of single-photon pair.
For example, we choose to use vacuum, two different intensities in $Z$ basis, and only one (weak) intensity in $X$ basis.  As we shall show,  both methods can improve the key rate. As shown below, given the same intensities, normally the experimental data would cause the fact that $\underline{s_{11}^Z}>\underline{s_{11}^X}$. We use different intensities in $Z$ basis
   and simply use $\underline{s_{11}^Z}$  for $s_{11}^{X}$, whenever we need $s_{11}^X$ n the all calculations. The differen intensities in $Z$ basis can be chosen based on the optimization of the final key. We only need {\em one} weak intensity in $X$ basis in order to estimate the upper bound of phase-flip rate of detected events by single-photon pairs.
   On the other hand,  since the observed bit-flip rate in $X$ basis is very large, the data from $X$ basis cannot be distilled into final key. We can also choose to use different {\em weak} intensities
   to find $\underline{s_{11}^X}$, and then use this as the fixed value for yield of single-photon pairs in both bases. In this method, we can choose {\em one} intensity in $Z$ basis  to optimize the key rate.

For clarity, lets recall the decoy-state MDI-QKD and denote some mathematical notations first.
In the protocol, we assume Alice (Bob) has three sources, $o_A,x_A,y_A$ ($o_B,x_B,y_B$) which can only emit three different states $\rho_{o_A}=|0\rangle\langle 0|, \rho_{x_A}, \rho_{y_A}$ ($\rho_{o_B}=|0\rangle\langle 0|, \rho_{x_B}, \rho_{y_B}$), respectively, in photon number space.
Suppose $\rho_{x_A}=\sum_{k} a_k \oprod{k}{k}$, $\rho_{y_A}=\sum_{k} a_k' |k\rangle\langle k|$, $\rho_{x_B}=\sum_{k} b_k |k\rangle\langle k|$, $\rho_{y_B}=\sum_{k} b_k' |k\rangle\langle k|$,

In the protocol, each time a pulse-pair (two-pulse state) is sent to the relay for detection. The relay is controlled by an UTP. The UTP will announce whether the pulse-pair has caused a successful event.
Those bits corresponding to successful events will be post-selected and further processed for the final key. Since real set-ups only use imperfect single-photon sources, we need the decoy-state method for security.

We assume Alice (Bob) has three sources, $o_A,x_A,y_A$ ($o_B,x_B,y_B$) which can only emit three different states $\rho_{o_A}=|0\rangle\langle 0|, \rho_{x_A}, \rho_{y_A}$ ($\rho_{o_B}=|0\rangle\langle 0|, \rho_{x_B}, \rho_{y_B}$), respectively, in photon number space.
Suppose
\begin{eqnarray}
\rho_{x_A}=\sum_{k} a_k |k\rangle\langle k|,&\quad& \rho_{y_A}=\sum_{k} a_k' |k\rangle\langle k|;\\
\rho_{x_B}=\sum_{k} b_k |k\rangle\langle k|,&\quad& \rho_{y_B}=\sum_{k} b_k' |k\rangle\langle k|,
\end{eqnarray}
and we request the states satisfy the following very important condition:
\begin{equation}\label{cond1}
\frac{a_k'}{a_k}\ge \frac{a_2'}{a_2}\ge \frac{a_1'}{a_1};\quad \frac{b_k'}{b_k}\ge \frac{b_2'}{b_2}\ge \frac{b_1'}{b_1},
\end{equation}
for $k\ge 2$.
The explicit formula for lower bound of $s_{11}$ with 3 different intensities of light was first given in Ref.\cite{wangPRA2013}. Very recently, a tighter bound was proposed~\cite{Wang2013}. Here we use the formula in Ref.\cite{Wang2013} for the lower bound value of single-photon pairs in $\omega$ basis, i.e. $s_{11}^{\omega},(\omega=X,Z)$ and the upper bound of the error rate $e_{11}^{\omega},(\omega=X,Z)$ with the following formulas
\begin{widetext}
  \begin{equation}\label{s11L}
    s_{11}^{\omega}\geq\underline{s}_{11}^{\omega}=\frac{(a_1 a'_2 b_1 b'_2-a'_1 a_2 b'_1 b_2)\tilde{S}_{xx}^{\omega}-b_1 b_2(a_1 a'_2-a'_1 a_2)\tilde{S}_{xy}^{\omega}-a_1 a_2 (b_1 b'_2-b'_1 b_2)\tilde{S}_{yx}^{\omega}}{a_1 b_1(a_1 a'_2-a'_1 a_2)(b_1 b'_2-b'_1 b_2)},
  \end{equation}
\end{widetext}
and
\begin{equation}\label{e11U}
  e_{11}^{\omega}\leq\overline{e}_{11}^{\omega}=\frac{\tilde{T}_{xx}^{\omega}}{a_1 b_1 \underline{s}_{11}^{\omega}},
\end{equation}
where $\tilde{S}_{xx}^{\omega}= S_{xx}^{\omega}-a_0 S_{0x}^{\omega}-b_0 S_{x0}^{\omega}+a_0 b_0 S_{00}, \tilde{S}_{xy}^{\omega}= S_{xy}^{\omega}-a_0 S_{0y}^{\omega}-b'_0 S_{x0}^{\omega}+a_0 b'_0 S_{00}, \tilde{S}_{yx}^{\omega}= S_{yx}^{\omega}-a'_0 S_{0x}^{\omega}-b_0 S_{y0}^{\omega}+a'_0 b_0 S_{00}, \tilde{S}_{yy}^{\omega}= S_{yy}^{\omega}-a'_0 S_{0y}^{\omega}-b'_0 S_{y0}^{\omega}+a'_0 b'_0 S_{00}$ and $\tilde{T}_{xx}^{\omega}= T_{xx}^{\omega}-a_0 T_{0x}^{\omega}-b_0 T_{x0}^{\omega}+a_0 b_0 T_{00}$, with $S_{\alpha\beta}^{\omega}, T_{\alpha\beta}^{\omega}$ are the probabilities of a bit or a wrong bit is produced whenever the source  $\alpha_A\beta_B$ is used in $\omega$ basis. We also have the relationship $T_{\alpha\beta}=S_{\alpha\beta}E_{\alpha\beta}$.

Consider those post-selected bits cased by source $x_A x_B$ in the $Z$ basis. After an error test, we know the bit-flip error rate of this set, say $T_{xx}^{Z}=E_{xx}^{Z}S_{xx}^{Z}$. We also need the phase-flip rate for the subset of bits which are caused by the two single-photon pulse, say $e_{11}^{ph}$, which is
asymptotically  equal to the flip rate of post-selected bits caused by a single photon in the $X$ basis, say $e_{11}^{X}$. Given this, we can now calculate the key rate. For example, for those post-selected bits caused by source $yy$, it is\cite{ind2}
\begin{equation}\label{KeyRate}
  R=a'_1 b'_1 s_{11}^{Z}[1-H(e_{11}^{X})]-f S_{yy}^{Z}H(E_{yy}^{Z}),
\end{equation}
where $f$ is the efficiency factor of the error correction method used.

As shown in \cite{wangPRA2013}, the exact values of yields of single-photon pairs must be equal in different bases. Suppose that at each side, horizontal and vertical polarizations have equal probability to be chosen. For all those single-photon pairs in the $Z$ basis, the state in polarization space is
\begin{equation}\label{OmegaHV}
  \frac{1}{4}\left(\Omega_{11}^{HH}+\Omega_{11}^{VV}+\Omega_{11}^{HV}+\Omega_{11}^{VH}\right)=\frac{1}{4}I,
\end{equation}
where $\Omega_{11}^{PQ}=\oprod{P}{P}\otimes \oprod{Q}{Q}$,$P,Q$ indicate the polarization which can be either $H$ or $V$. On the other hand, for all those two single-photon pulse pairs prepared in the $X$ basis, if the $\pi/4$ and $3\pi/4$ polarizations are chosen with equal probability, one can easily find that the density matrix of these single-photon pairs is also $I/4$. Therefore, we conclude
\begin{equation}\label{s11ZX}
  s_{11}^{Z}=s_{11}^{X}.
\end{equation}

If we implement the decoy-state method for different bases separately, with the known values $S_{\alpha\beta}$, we can calculate the values of $\underline{s_{11}^{X}}$ and $\underline{s_{11}^{Z}}$ by Eq.(\ref{s11L}). Usually, these two estimated values are not equal to each other. Actually, we can find out that $\underline{s_{11}^{Z}}\geq \underline{s_{11}^{X}}$ in our simulations for coherent-state source  if we use same intensities in each bases. Taking the fact that $s_{11}^{X}=s_{11}^{Z}$ into consideration, we can substitute
$\underline{s_{11}^X}$ with $\underline{s_{11}^Z}$ and have a more tight upper bound of $e_{11}^{X}$ such that
\begin{equation}\label{e11XZ}
  \overline{e}_{11,Z}^{X}=\frac{\tilde{T}_{xx}^{\omega}}{a_1 b_1 \underline{s_{11}^{Z}}}\leq \overline{e}_{11}^{X},
\end{equation}
where $\underline{s_{11}^{Z}}$ is given in Eq.(\ref{s11L}) and $\overline{e}_{11}^{X}$ is defined in Eq.(\ref{e11U}). Then we can use the following formula to calculate the key rate
\begin{equation}\label{KeyRateZ}
  R_{Z}=a'_1 b'_1 \underline{s_{11}^{Z}}[1-H(\overline{e}_{11,Z}^{X})]-f S_{yy}^{Z}H(E_{yy}^{Z}),
\end{equation}
where $\underline{s_{11}^{Z}}$ is given in Eq.(\ref{s11L}) and $\overline{e}_{11,Z}^{X}$ is defined in Eq.(\ref{e11XZ}). With Eq.(\ref{KeyRateZ}), we only need the weaker decoy-state pulse in $X$ basis. That is to say, in this situation, Alice and Bob only need sources $x_A$ and $x_B$ in $X$ basis. Given the intensities of sources $x_A,x_B$ in $X$ and $Z$ bases, we can optimize the intensity of sources $y_A,y_B$ in $Z$ basis by maximizing the key rate $R_{Z}$. In Eq.(\ref{KeyRateZ}), $\underline{s_{11}^{Z}}$, $\overline{e}_{11,Z}^{X}$ and $S_{yy}^{Z}H(E_{yy}^{Z})$ are all dependent on the intensity. There is a balance between $\underline{s_{11}^Z}$ and $S_{y_Ay_B}$. These restrictions make the key rate value being limited: a bigger intensity will produce a larger $S_{yy}$, but will also lead to a smaller $\underline{s_{11}^Z}$. can not be increased remarkably. Given this fact, we can also consider to replace $\underline{s_{11}^{Z}}$ by $\underline{s_{11}^{X}}$ to get another formula to calculate the key rate
\begin{equation}\label{KeyRateX}
  R_{X}=a'_1 b'_1 \underline{s_{11}^{X}}[1-H(\overline{e}_{11}^{X})]-f S_{yy}^{Z}H(E_{yy}^{Z}),
\end{equation}
where $\underline{s_{11}^{Z}}$ is given in Eq.(\ref{s11L}) and $\overline{e}_{11}^{X}$ is defined in Eq.(\ref{e11U}).
If we want to use this formula, we need different intensities in $X$ basis to figure out $\underline{s_{11}^X}$, while only need one intensity (signal pulse) in $Z$ basis.  Since the pulses in $X$ basis cannot be used to generate the final key, here we can choose smaller values of intensities so as to produce a bigger value of $\underline{s_{11}^X}$. We shall also regard this as the lower bound for $s_{11}^Z$. Now that the lower bound of $s_{11}^Z$ is determined by the light in $X$ basis already, in optimizing the key rate, we only need to choose an optimal intensity for source $y_A$, $y_B$ so that $S_{yy}^Z$ is maximized. Note that here in choosing the optimum intensity in $Z$ basis, there is no balance between $S_{yy}^Z$ and $\underline{s_{11}^Z}$, as the latter one has been determined by pulses in $X$ basis already, therefore we only need to optimize $S_{yy}^Z$. This gives us larger freedom in choosing the intensity and hence offers more chances for a higher key rate.



\begin{figure}
  \includegraphics[width=240pt]{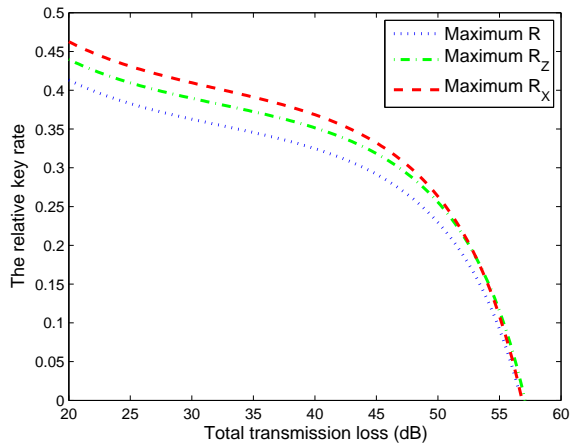}\\
  \caption{(Color online) The optimaized relative key rate versus the total channel transmission loss using 3-intensity decoy state MDI-QKD. Values are the ratio of key rate to the key rate using infinite intensities.}\label{rKeyRate}
\end{figure}

\begin{figure}
  \includegraphics[width=240pt]{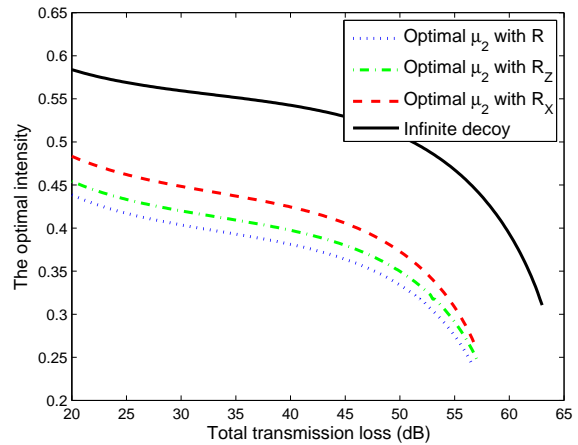}\\
  \caption{(Color online) The optimal intensities versus the total channel transmission loss using 3-intensity decoy state MDI-QKD.}\label{OptRmu}
\end{figure}

\begin{table}
\caption{\label{tabPara}List of experimental parameters used in numerical simulations: $e_0$ is the error rate of background, $e_d$ is the misalignment-error probability; $p_d$ is the dark count rate per detector; $f$ is the error correction inefficiency.}
\begin{ruledtabular}
\begin{tabular}{cccc}
  $e_0$ & $e_d$ & $p_d$ & $f$ \\
  \hline
  0.5 & 1.5\% & $3.0\times 10^{-6}$ & 1.16\\
\end{tabular}
\end{ruledtabular}
\end{table}

Now, we present some numerical simulations to comparing our results with the existing results~\cite{Wang2013,LiangPRA2013}. Below for simplicity, we suppose that Alice and Bob use the coherent-state sources.  Here, we denote Alice's sources $\{0_A,x_A,y_A\}$ by their intensities $\{\mu_0,\mu_1,\mu_2\}$ and Bob's sources $\{0_B,x_B,y_B\}$ by their intensities $\{\nu_0,\nu_1,\nu_2\}$ respectively.  The UTP locates in the middle of Alice and Bob, and the UTP's detectors are identical, i.e., they have the same dark count rate and detection efficiency, and their detection efficiency does not depend on the incoming signals. We shall estimate what values would be probably observed for the gains and error rates in the normal cases by the linear models as in~\cite{ind2,LiangPRA2013}:
\begin{eqnarray*}
  |n\rangle\langle n| = \sum_{k=0}^n C_n^k \xi^k (1-\xi)^{n-k}|k\rangle\langle k|
\end{eqnarray*}
where $\xi^k$ is the transmittance for a distance from Alice to the UTB.  For fair comparison, we use the same parameter values used in~\cite{ind2,LiangPRA2013} for our numerical evaluation, which follow the experiment reported in~\cite{UrsinNP2007}. For simplicity, we shall put the detection efficiency to the overall transmittance $\eta=\xi^2 \zeta$. We assume all detectors have the same detection efficiency $\zeta$ and dark count rate $p_d$. The values of these parameters are presented in Table~\ref{tabPara}. With this, the total gains $S_{\mu_i,\nu_j}^{\omega},(\omega=X,Z)$ and error rates $S_{\mu_i,\nu_j}^{\omega}E_{\mu_i,\nu_j}^{\omega},(\omega=X,Z)$ of Alice's intensity $\mu_i (i=0,1,2)$ and Bob's intensity $\nu_j (j=0,1,2)$ can be calculated.  By using these values, we can estimate the key rate with Eq.(\ref{KeyRate}), Eq.(\ref{KeyRateZ}) and Eq.(\ref{KeyRateX}) in Fig.\ref{KeyRate}, which shows that our methods are more tightly than the pre-existed result. In order to see more clearly, in Fig.\ref{rKeyRate}, we plot the relative value of the key rate to the result obtained with the infinite decoy-state method. We can observe that our results are closer to the asymptotic limit of the infinite decoy-state method than the pre-existed results. Note that, in the actual case, the advantage of our method will be even larger than presented in the figure. In the actual case, the total number of pulses is finite. In our protocol, we use fewer intensities than the existing one does. Therefore the the probability of intensity mismatch becomes smaller. In these figures, the blue dotted line is obtained by Eq.(\ref{KeyRate}), the green dash-dot line is obtained by Eq.(\ref{KeyRateZ}), the red dashed line is obtained by Eq.(\ref{KeyRateX}), and the black solid line is obtained by the infinite decoy-state method. In the simulation, the densities used by Alice and Bob are assigned to $\mu_1=\nu_1=0.1$, $\mu_2=\nu_2=0.15$.  The optimal densities with maximizing the key rate versus the total channel transmission loss is given in Fig.\ref{OptRmu} with the blue dotted line, the green dash-dot line and the red dashed line corresponding to the key rate given by Eq.(\ref{KeyRate}), Eq.(\ref{KeyRateX}), and Eq.(\ref{KeyRateZ}) respectively.

In summary, we have shown that the decoy-state MDI QKD can be done with different intensities in only one basis. Our method has the a number of advantages: We use fewer intensities and this may simplify the implementation; also our method reduces the probability of intensity mismatch, this will definitely improve the key rate. Even we don't consider the factor of intensity mismatch and only  consider the key rate from those signal pulses, our method can still offer a higher key rate because we can choose to use the bigger  value between $\underline{s_{11}^Z}, \underline{s_{11}^X}$.

{\bf Acknowledgement:}
We acknowledge
the support from the 10000-Plan of Shandong province,
the National High-Tech Program of China Grants
No. 2011AA010800 and No. 2011AA010803 and NSFC
Grants No. 11174177 and No. 60725416.


\end{document}